\documentclass[
 reprint,
 preprintnumbers,
 amsmath,
 amssymb,
 aps,
 prstab,
 floatfix,
]{revtex4-1}

\usepackage{graphicx}			
\usepackage{verbatim}			
\usepackage{wasysym}			
\usepackage{mathtools}			
\usepackage[makeroom]{cancel}		
\usepackage{dcolumn}			
\usepackage{hyperref}			
\usepackage{color}			

\newcommand{\pd}{\partial}				
\newcommand{\dd}{\mathrm{d}}				

\newcommand{\omsc}{\omega^\mathrm{sc}}
\newcommand{\omsci}{\omega^\mathrm{sc}_i}

\newcommand{\omxi}{\omega_{x i}}
\newcommand{\domxi}{\delta \omega_{x i}}

\newcommand{\omx}{\omega_{x}}
\newcommand{\omc}{\omega^\mathrm{c}}
\newcommand{\oms}{\omega_\mathrm{s}}

\newcommand{\omi}{\omega_{i}}
\newcommand{\domi}{\delta \omega_{i}}
\newcommand{\domo}{\delta \omega_{0}}

\newcommand{\xbar}{\bar{x}}

\newcommand{\Dom}{\Delta \omega}

\begin{document}

\title{Space Charge Effects for Collective Instabilities in Circular Machines}
\author{Alexey Burov}
\email{burov@fnal.gov}
\affiliation{Fermilab, PO Box 500, Batavia, IL 60510-5011}
\date{\today}

\begin{abstract}

Coulomb fields of charged particle beams in circular machines determine, together with wake fields, modes of the collective beam oscillations, both for transverse and longitudinal degrees of freedom. Recent progress in these two areas of beam dynamics is discussed.    
 
\end{abstract}


\maketitle

 



\section{\label{sec:Int}Introduction}

This paper discusses recent progress in the area of collective linear dynamics of charged particle beams, when their space charge (SC) is important; both transverse and longitudinal planes are considered. Since such a task cannot be free from some subjectivity and arbitrariness, I beg a pardon for the inevitable omission of some valuable results. 

More than fifty years ago, the first significant publication was presented on transverse collective instabilities of space-charge-dominated beams in circular machines; it was a preprint of G.~Merle and D.~M\"{o}hl ``The stabilizing influence of nonlinear space charge on transverse coherent oscillations''~\cite{Merle:1969zz}. A relatively simple equation of motion was suggested there as something obvious. Although it was, strictly speaking, neither obvious nor necessarily justified, it played and continues to play an extraordinarily important role. This equation implicitly relied on rigidity of the beam slices; four decades later it was shown  that this approximation requires strong SC, i. e. the SC tune shift to exceed the lattice tune shifts and the synchrotron tune~\cite{BurLebPRAB2009}. Soon after that, the concept of strong SC was used for bunched beams~\cite{burov2009head}, including computation of Landau damping (LD) with space charge and octupoles taken into account. It was also demonstrated there, that SC can help to avoid the transverse mode coupling instability (TMCI). Paradoxical features of this stabilization were realized few years ago, leading to discovery of the transverse convective instabilities~\cite{BurovConvectivePRAB2019}, with exponential dependence of the head-to-tail amplification on the SC tune shift. These and related ideas are discussed in Sec.~\ref{SecTrans}, devoted to the transverse collective oscillations.  

Longitudinal modes are discussed in Sec.~\ref{SecLong}, for the bunched beams only, since the coasting beam theory seems to be well established long ago, see e.g. Refs.~\cite{chao1993physics, ng2006physics}. For bunched beams, the key intensity limitation is associated with the loss of LD, when a single-bunch impedance drives a collective mode outside the incoherent spectrum, making the beam defenseless against any coupled-bunch instability, no matter how weak it might be. First time this issue was noted and analyzed in this context by F.~Sacherer almost half a century ago~\cite{SachererLLD}; a certain threshold for SC impedance below transition was estimated. Later this problem was addressed by other authors, who got the same result, up to a numerical factor; the smallest threshold was reported in Ref.~\cite{burov2012dancing}; it was about three times below the Sacherer's one. These discrepancies were related to the model imperfections and to high sensitivity of the result on the distribution function. However, a recently published article of Karpov et al.~\cite{KarpovPRAB2020} has shown that all those results are incorrect; for the pure SC impedance the LLD intensity threshold is equal to zero. Soon after that, the entire collective spectrum for SC impedance below transition was described in Ref.~\cite{BurovPRAB2021WeakSC}, with generalization to any multipolarity and to other impedances.The single-bunch eigensystem problem was reduced to a parameter-less equation; also the coupled-bunch growth rates were analytically derived. These and related ideas are considered in Sec.~\ref{SecLong}.

\section{\label{SecTrans}Transverse Oscillations}

\subsection{\label{SecCoast} Coasting Beams}

To analyze the beam stability with SC, a linear equation of motion was suggested by G.~Merle and D.~M\"{o}hl in 1969~\cite{Merle:1969zz}:
\begin{equation}
\frac{d^2 x_i}{dt^2} + \omxi^2 x_i + 2 \omx \omsci (x_i -\bar{x}) +  2 \omx \omc \bar{x} =0 \,.
\label{MMEq}
\end{equation}
Here $x_i=x_i(t)$ is a transverse offset of a particle $i$, $\omxi$  is the betatron frequency of the particle $i$, $\omx$ is the average betatron frequency, $\xbar=\xbar(t)$ is an average offset of that beam {\it slice} where the particle $i$ is located at the given moment of time $t$, $\omsci <0$ is the SC frequency shift of the particle, and $\omc$ is the coherent frequency shift parameter, proportional to the ring impedance.  The full time derivative $d/dt$ is expressed through the partial ones, $d/dt = \pd/\pd t + \omi \pd/\pd \theta$, where $\omi$ is the revolution frequency of the particle, and $\theta =s/R$ is the azimuthal angle, with $s$ as the conventional longitudinal coordinate and $R$ as the average ring radius. The term `slice' refers to the group of beam particles which Coulomb fields affect the given particle number $i$, i.e. the particles with positions somewhere between $s_i -a/\gamma$ and $s_i +a/\gamma$, where $a$ is the beam transverse size and $\gamma$ is the Lorentz factor.   

Equation~(\ref{MMEq}) implies two important things. 

First, it implies that $x_i$ relates to the driven part of the single-particle oscillations, excited by the collective motion of the slices' centroids $\xbar$, while constant amplitudes of free oscillations determine the space charge frequency shifts. That is why the offset $x_i$ is of the order of centroid offsets, $x_i \sim \xbar$, so it can be considered infinitesimally small, while the incoherent amplitudes are of the order of the beam transverse size.  

Second, this equation assumes that each beam slice oscillates as a rigid body, allowing for a representation of the SC force in the simple way it is done there. Because of this assumption, Merle-M\"{o}hl approach is sometimes addressed as the {\it rigid-slice} or {\it frozen-field} model. Possible incorrectness of this assumption, as well as its very existence, was realized much later, when some strange features of Eq.~(\ref{MMEq}) were discovered.  

For a coasting beam, eigenfunctions of Eq.~(\ref{MMEq}) have the form 
\begin{equation}
x_i\,, \xbar \propto \exp\left[-i\,(\omx + n\omega_0 + \omega)t + in \theta \right]\,
\label{eigen1}
\end{equation}
where $n$ is an arbitrary integer, $\omega_0$ is the average revolution frequency, and $\omega$ is a frequency shift of the mode $n$. Substitution of this form into Eq.~(\ref{MMEq}) yields for the complex amplitudes
\begin{equation}
x_i = \xbar \frac{\omc - \omsci}{\omega - \omsci -\domxi -n\domi}\,,
\label{eigen2}
\end{equation}
with the lattice frequency shifts $\domxi=\omxi -\omx$ and $\domi=\omi-\omega_0$. 
Note that without lattice frequency spread, $\domxi=\domi=0$, there is always the rigid-bunch solution, $x_i=\xbar$, with $\omega=\omc$, independently of the SC tune shifts $\omsci$. This important physical property of Eq.~(\ref{MMEq}) is a consequence of its SC representation by means of the term $\propto x_i - \xbar$. In fact, the Merle-M\"{o}hl equation is the only possible linear dynamic equation consistent with the given incoherent spectrum, its lattice and SC parts, which represents the coherent SC term by means of $\propto \xbar$ term only, preserving the rigid-bunch mode for zero lattice tune spread, as it must be from the first principles.  

By averaging over the particles, writing the sums as the phase space integrals with the distribution function, one gets the dispersion relation, i.e. the equation for the sought-after eigenfrequency $\omega$, 
\begin{equation}
1 = - \int d\Gamma \, J_x \frac{\pd f}{\pd J_x} \frac{\omc - \omsci}{\omega - \omsci -\domxi -n\domi +i o}
\label{DispEq}
\end{equation}
Here $f=f(J_x, J_y, \delta p/p)$ is the unperturbed distribution density as a function of transverse actions $J_{x,y}$ and the relative momentum offset $\delta p/p$, normalized to unity, $\int d\Gamma f =1$, where $d\Gamma = dJ_x\,dJ_y\, d\delta p/p$; the single-particle subscript $i$ has to be understood as indication to related functional dependences, i.e. $\domxi \rightarrow \delta \omega_x(J_x, J_y, \delta p/p)$, etc. To get Eq.~(\ref{DispEq}) from Eq.~(\ref{eigen2}), the Hereward rule~\cite{hereward1969landau} was used,
$$\sum_i (...) \rightarrow - \int d\Gamma \, J_x\, \pd f/\pd J_x \,(...) \,,$$  
and the Landau rule of going around the pole is explicitly marked, $\omega \rightarrow \omega + i o$, where $o$ is an infinitesimally small positive number. 

It is straightforward to see from the dispersion relation~(\ref{DispEq}) that without lattice frequency spreads, at $\domxi=\domi=0$, the eigenfrequency $\omega=\omc$, independently of the SC tune shift. Thus, even if the phase space density of the resonant particles were not zero, i.e. there were particles with the same tune as the coherent mode, still there would be no Landau damping (LD), irrespectively to nonlinearity of SC distribution.

The dispersion relation in the form~(\ref{DispEq}) was first derived by D.~M\"{o}hl and H.~Sch\"{o}nauer in 1974~\cite{Mohl:1974vj}, not in the original preprint of Merle and M\"{o}hl. In the latter, some mathematical mistakes were adopted, so the dispersion relation was derived incorrectly. Due to this, it was mistakenly concluded there that SC nonlinearities may contribute to Landau damping of coasting beams even without the lattice frequency spread. This mistake was later repeated in Ref.~\cite{reich1972transverse} and corrected by M\"{o}hl and Sch\"{o}nauer~\cite{Mohl:1974vj}. That is why it seems fair to call Eq.~(\ref{MMEq}) Merle-M\"{o}hl equation of motion and Eq.~(\ref{DispEq}) M\"{o}hl-Sch\"{o}nauer dispersion relation.


After the simplest case of no lattice frequency spreads, the next by simplicity is a two-stream beam, $\domi=\pm \domo$. For the Kapchinsky-Vladimirsky transverse distribution with a constant SC frequency shift, $\omsci=\omsc$, Eq.~(\ref{DispEq}) yields the following spectrum: 
\begin{equation}
\omega=\frac{\omc + \omsc}{2} \pm \sqrt{\frac{(\omc - \omsc)^2}{4} + n^2 \domo^2}
\label{2stream} 
\end{equation}
The instability is driven by the coherent tune shift $\omc$, so the most unstable modes are positive ones, i.e. those associated with the sign $+$ in Eq.~(\ref{2stream}). A more detailed analysis shows that for them the two streams of the beam oscillate approximately in phase, so their wakes add to each other. For strong SC, $|\omsc| \gg \max(|n|\domo,\, |\omc|)$, the spectrum of positive modes reduces to
\begin{equation}
\omega= \omc + n^2 \domo^2/\omsc \,, 
\label{2stream2} 
\end{equation}
being far away from the incoherent spectrum localized around $\omsc$. Thus, the negative modes not only are barely excited by the wake, but also stay close to the incoherent spectrum, so their stability would be provided automatically when the positive modes are stable. 

A very general method of analysis of integral dispersion relations like Eq.~(\ref{DispEq}) was presented in Ref.~\cite{Ruggiero:1968vf}. The idea was to reverse the problem: instead of finding the eigenfrequency $\omega$ when the coherent tune shift $\omc$ is given, let us do the opposite, find the coherent tune shift $\omc$ for a given eigenfrequency $\omega$ for the same dispersion relation. If to run the eigenfrequency along the real axis, the corresponding coherent tune shift will follow a certain line in its complex plane, the conformal map of the real axis in the complex plane of $\omega$ to the complex plane of $\omc$. This line, $\omc(\omega)$, is conventionally called the stability diagram; the beam is stable if and only if its actual coherent tune shift is below the diagram. 

Certain historical investigations~\cite{VaccaroDiagram} convinced the author of this paper that it would be fair to call the stability diagram the Vaccaro diagram (VD), by name of Vittorio Vaccaro, who found how Nyquist's stability plots can be modified to become an effective tool for the collective beam dynamics. With M\"{o}hl-Sch\"{o}nauer dispersion relation~(\ref{DispEq}), VD is determined by the beam distribution function $f$, the lattice frequency shifts and SC tune shifts, being independent of wakes.

In 2001, M.~Blaskiewicz suggested an original method of solving the Jeans-Vlasov equation with SC and lattice tune spread~\cite{Blaskiewicz:PRAB2001}; with respect to the equation naming, see Ref.~\cite{henon1982vlasov}. The method was free from hidden assumptions of Merle-M\"{o}hl approach, being, arguably, more complicated and less transparent in the computations. The method allowed to make conclusions regarding SC effects for the stability diagram. In case of the chromatic lattice tune spread, the diagram of a Gaussian beam essentially shifted to the left by one half of the maximal SC tune shift. In case of the octupole nonlinearity M.~Blaskiewicz found that focusing octupoles are much more beneficial for LD than defocusing ones, confirming the same conclusion by D.~M\"{o}hl~\cite{Mohl:1995gb}.   

The first analytical attempt to build VD for Eq.~(\ref{DispEq}) was presented in 2004 by E.~Metral and F.~Ruggiero~\cite{Metral:2004qw}. Namely, they suggested a solution of the dispersion relation with SC and octupolar nonlinearity, where, instead of the coasting beam term $n\domi$, the bunched beam term $k \oms$ was put, with $\oms$ as the synchrotron frequency and $k$ as the head-tail mode number. Such extension of the coasting beam theory to the bunched case was, however, left unexplained both in the paper itself and the references it suggested for that matter, including Ref.~\cite{Mohl:1995gb}. As it became more clear later, this extension works reasonably well only when the SC term could be safely omitted. However, the actual merit of the paper was not in its applicability to bunched beams with SC, but in its analytical building of VD for coasting beams with octupoles, nonlinear SC, and insignificant revolution frequency spread, $n\domi =0$. It was confirmed, in particular, that the octupole sign becomes crucial for strong SC; namely, the focusing octupoles are much preferable. The reason is that the octupoles affect mostly the tail particles, so the collective frequency is barely touched by them. Landau damping requires resonant particles, i.e. those which individual tunes are the same as the collective one. Space charge moves the incoherent tunes down, and does almost nothing for the collective tunes, thus killing LD. Thus, to restore the latter, the incoherent tunes have to be moved up to provide higher population of the resonant particles, so the octupoles have to be focusing. It was also shown in this reference that SC can be beneficial if it shifts VD on top of the coherent tune, which would be outside (on the left) of VD without space charge. For very strong SC, it meant that it is detrimental since it shifts the stability diagram far on the left.

Among multiple reasonable features, Vaccaro diagrams of Ref.~\cite{Metral:2004qw} showed a strange one: for defocusing octupoles, there was a kink point of the curve at the real axis, which prevented the line from going to the lower half-plane, $\Im \omc <0$. The kink point looked strange, since VD should be analytical by the definition. 

In 2006, D. Pestrikov published an article~\cite{Pestrikov:2006} where a similar problem was solved, but instead of the kink point, the diagram smoothly continued to the lower half-plain, thus demonstrating Landau antidamping. Later that year Landau antidamping was confirmed by K.Y. Ng for the same model as Metral and Ruggiero proposed~\cite{MetralNg2006}. On the ground of these findings, the kink point of Ref.~\cite{Metral:2004qw} was dismissed as a mistake of the sign. Due to this, however, another problem appeared: at certain conditions, a Gaussian-like beam with time-independent Hamiltonian started looking unstable even when the coherent tune shift $\omc$ suggested a decay of the mode. Next year Pestrikov published another paper~\cite{Pestrikov:2007}, presenting ``a self-consistent model'' which showed no antidamping, contrary to the Merle-M\"{o}hl model; he expressed doubt in the validity of the latter. 

This doubt was enhanced to a stronger claim by V.~Kornilov, O.~Boine-Frankenheim, and I.~Hofmann in their publication of 2008~\cite{Kornilov:2008zz}. First, they confirmed that VD of Eq.~(\ref{DispEq}) indeed yields Landau antidamping for defocusing octupoles. Second, they supported this confirmation by macroparticle simulations within the {\it frozen field} model, equivalent to Merle-M\"{o}hl approach. Third, they ran self-consistent macroparticle simulations for the same conditions, and saw no antidamping. From this, they concluded that ``antidamping can be related to the non-self-consistent treatment of nonlinear space charge in the simulations and also in the dispersion relation.'' 

At that stage, several issues remained unresolved for coasting beams. First, it was not clear if Landau antidamping is ever possible for Gaussian-like beams with SC, octupoles and chromaticity. Second, with evidence of incorrectness of Merle-M\"{o}hl analytical approach at certain cases, it was not clear if their equation could be ever used at all, and under what conditions. Third, no analytical formulas for the instability thresholds were yet obtained. These issues were addressed in Ref.~\cite{BurLebPRAB2009}.

A possibility of Landau antidamping was denied there as contradicting to the Second Law of Thermodynamics. Indeed, a beam with real coherent tune shift $\omc$, corresponding to imaginary transverse impedance, i.e. to zero energy losses, can be described by energy-preserving time-independent Hamiltonian, so the growing coherent oscillations might take energy from the incoherent degrees of freedom only. For a Gaussian beam it would mean a perpetuum mobile of the second kind, forbidden by the Second Law. Landau antidamping, demonstrated for some parameters by Merle-M\"{o}hl dynamic system~(\ref{MMEq}), is caused by the non-Hamiltonian character of its SC term. Specifically, the term $\propto \omsci \xbar $ is non-Hamiltonian unless all the SC frequency shifts are identical within the beam slice. Having said that, it is important to stress that the Merle-M\"{o}hl equation of motion with real coherent tune shift, $\Im \omc=0$, may lead to Landau antidamping only if the incoherent spectrum $\omsci + \domxi$ reaches a local maximum in the action space, which may happen for a defocusing octupole. Although the equation is not Hamiltonian, for monotonic spectra $\omsci + \domxi$ all its van Kampen eigenfrequencies with real coherent tune shift are real as well, no unphysical dissipation is introduced.   

How reliable is Eq.~(\ref{MMEq}) for LD computation for the monotonic spectra?  When the SC tune shifts depend on the transverse actions, as they normally are, the defect of the model still should not play a role, if the slices were sufficiently rigid in their transverse oscillations. In this case only the tail particles would be responsible for LD, so the energy transfer to them could be reasonably approximated with the rigid core model. To see when the core is really rigid, note that if the lattice tune shifts are small with respect to the tune separation,  
\begin{equation}
|\domxi + n\domi| \ll |\omc - \omsci|\,,
\label{SSCcond1}
\end{equation}
the particles move together with the related centroids, $x_i \approx \xbar$, as it follows from Eq.~(\ref{eigen2}). Thus, if the SC is so strong that this condition is satisfied for the majority of particles, the slices oscillate almost without distortions, since almost all the particles oscillate almost identically to their centroids; so the rigid-slice approximation is justified. Luckily, for many low- and medium-energy machines, typical SC tune shifts are much larger than the imaginary part of the coherent tune shifts, $|\omsci | \gg \Im \omc$, so stabilization is achieved at such a small lattice nonlinearity that Eq.~(\ref{SSCcond1}) is satisfied, justifying Merle-M\"{o}hl equation. In this case of strong SC, instability thresholds were explicitly found in Ref.~\cite{BurLebPRAB2009} for Gaussian beam, both for octupolar and chromatic frequency spreads. Recently, this method was extended to electron lenses; Landau damping rate introduced by a Gaussian e-lens for a coasting beam with SC was analytically estimated and presented in Ref.~\cite{LebECooler}.

\subsection{\label{SecBunch} Bunched Beams}

The coherent spectrum of bunched beam with SC was presented for the first time by M.~Blaskiewicz in 1998~\cite{blaskiewicz1998fast} within a simple model of an air-bag bunch in a square potential well. For a delta-wake, the eigenfrequencies were found to be the same as Eq.~(\ref{2stream}) for the two-stream coasting beam, with the substitution $n \domo \rightarrow k \oms$, where $k=0, 1, 2,...$ is the mode counter and $\oms$ is the synchrotron frequency. 
A new and rather surprising mathematical result of M.~Blaskiewicz~\cite{blaskiewicz1998fast} showed suppression of the transverse mode coupling instability (TMCI) by SC; the wake threshold was demonstrated to grow with SC tune shift, linearly at the strong SC limit, $|\omsc | \gg \oms$. This result was obtained for exponential wakes and the ABS model (Air-Bag, Square-well), so a question was raised about the sensitivity of this unexpected result to the details of the wake, potential well and bunch distribution. Also, it was not clear if there was any limit to this growth of the instability threshold. An explanation of this growth at moderate SC was suggested to the author by V.~Danilov~\cite{SlavaComm1998} and reproduced in Ref.~\cite{Ng:1999fy}. Without SC, TMCI typically results from crossing of the head-tail mode $0$, shifted down by the wake, and the mode $-1$, not shifted as much. Space charge, on the contrary, does not influence the mode $0$ and shifts down the mode $-1$, thus moving their coupling point to higher intensity.      

In the year of 2009, when it was understood that Merle-M\"{o}hl approach of rigid slices is justified for sufficiently strong space charge, it was applied to bunched beams by the author~\cite{burov2009head}. Under the condition of SC tune shift being much stronger than all other tune shifts and spreads, as well as the synchrotron tune (strong space charge, SSC), an ordinary linear integro-differential equation was derived for the bunch modes for an arbitrary potential well, driving and detuning wakes, longitudinal and transverse bunch distribution functions. Later that same year V.~Balbekov published a paper~\cite{Balbekov:2009PRAB} with an alternative derivation of the SSC equation, whose result differed from mine. After checking his derivation and rechecking mine, I found an algebraic error in my calculations, and derived my ultimate form of the SSC mode equation, which agreed with Balbekov's result, suggesting a slightly more compact form in the erratum~\cite{Burov2009PRABErratum}, 
\begin{equation}
i\frac{\pd \xbar}{\pd t} + \frac{1}{\omsc} \frac{\pd }{\pd s} \left( u^2 \frac{\pd \xbar}{\pd s} \right)= \mathbb{W} \xbar + \mathbb{D} \xbar \,.
\label{MySSCEq}
\end{equation}
Here $\omsc = \omsc(s)$ is the SC frequency shift averaged over the transverse actions at every position $s$, $u^2=u^2(s)$ is the local rms spread of the longitudinal velocities, $u^2 \equiv \left < R^2 \domi^2 \right>$, while $\mathbb{W}$ and $\mathbb{D}$ are conventional driving and detuning wake linear integral operators~\cite{BurovDanilovPRL1999}; for more details see~\cite{Burov2009PRABErratum}. The equation is complemented by zero-derivative boundary conditions, $\pd \xbar /\pd s =0$ at the bunch edges or at $s=\pm \infty$. For the eigenfunctions, the time derivative has to be substituted by the sought-for eigenfrequency $\nu$, i.e. $i \pd /\pd t \rightarrow \nu$. Without wakes, this equation leads to the Blaskiewicz-type collective spectrum, $\nu_k \simeq k^2 \oms^2/\omsc$. The mathematical elegance of Eq.~(\ref{MySSCEq}) has its price: missing is the Landau damping, which required additional ideas and computations.  

Analytical estimations for LD at SSC were also suggested in Refs.~\cite{burov2009head, Burov2009PRABErratum} for weak head-tail cases, when the wake does not influence the eigenfunction much. Contrary to coasting beams, it was found that there is an intrinsic LD, caused by the longitudinal variation of the SC tune shift only, even without any lattice tune spreads. The physical mechanism of the dissipation was associated with a break of the slice rigidity at the bunch edges, where the SC is not strong any more. The slice softening at the bunch edges opens a way for the energy transfer to the incoherent degrees of freedom. According to the related estimation, the intrinsic LD rate $\Lambda_k$ at SSC was found to be a steep function of the SC parameter $q \equiv \omsc/\oms$ and the positive mode number $k$, $\Lambda_k \simeq k^4 \oms/q^3$; the SSC assumes $q \gg 2k$.  Six years later these analytical results for SSC eigenfunctions and LD rates were fully confirmed in Synergia macroparticle simulations by A.~Macridin et al.~\cite{MacridinPRAB2015}, where the intrinsic LD rates were shown to have their maxima at $q \simeq 2k$. A more subtle case of {\it parametric Landau damping} was treated by A.~Macridin et al. in Ref.~\cite{MacridinPRAB2018} by means of analytical modeling and macroparticle simulations. Analytical estimations of octupoles-related LD suggested in Refs.~\cite{burov2009head, Burov2009PRABErratum} are still waiting for at least numerical verifications; nothing yet has been published in that matter. Octupoles, however, are rather inefficient for LD, which requires significant nonlinearity inside the beam, not far outside, as octupoles provide. That is why a better instrument for LD is an electron lens, at least as thin as the beam. Such e-lenses are able to provide LD without deterioration of the dynamic aperture, as it was pointed out by V.~Shiltsev et al.~\cite{ShilAlexBurValPRL2017}. Estimations of e-lens-caused LD rates for bunches with SC were suggested by Y.~Alexahin, A.~Burov and V.~Shiltsev in 2017~\cite{AlexahinBurShil2017rep}.    

With the wake taken into account, the Blaskiewicz' result of linear growth of the TMCI wake threshold was confirmed in a series of publications, see Refs.~\cite{ZolkinBurovPandeyPRAB2018, BalbekovPRAB2019} and references therein. A hidden obstacle with this problem, sometimes caused  misleading results, was realized by V.~Balbekov~\cite{balbekov2017transverse}, who showed that convergence of the expansion of the sought-for eigenfunction over the zero-wake basis degrades with SC, requiring more and more terms for higher SC parameter $q$. The physical reason of the convergence worsening was recently found by the author of this paper; it is associated with the head-to-tail amplification, or the {\it convective instabilities} driven by wakes at SSC. When eigenfunctions are significantly amplified, their expansion over any even basis cannot be of a good convergence. Clearly manifest subsiding of the instability with SC was presented in the two-particle model of Ref.~\cite{ChinTwoPartPRAB2016}. 

With the theoretical proof of TMCI vanishing with SC, two problems became rather obvious, one experimental and the other theoretical. The former consisted in a reasonable agreement of the transverse instability at CERN SPS with no-SC theory, while SC tune shift was very strong there, especially with the old lattice~\cite{Bartosik:2013qji, MetralFNAL2018}. The latter problem was related to the linear growth of the wake threshold with SC. Due to this feature, the bunch should be stable up to infinite intensity, as soon as its emittance is low enough, which did not sound as a reasonable statement. The resolution of both problems was presented by the author few years ago~\cite{BurovConvectivePRAB2019}. The main idea, already mentioned above, was that, while moving out TMCI, SC moves another instability in its place, a convective one. Contrary to TMCI, which is an absolute instability, i.e. has nonzero growth rate, the convective instabilities grow not in time, but in space, from head to tail~\cite{LifPitKin}. This head-to-tail amplification increases exponentially with bunch intensity, resulting in one or another physical limit, set by lattice nonlinearity, beam loss or feedback. When the amplification is large, even a tiny feedback from tail to head may be sufficient to close the loop and turn the convective instability into an {\it absolute-convective} one, like those with a microphone close to its loudspeaker. Such a feedback may be presented with a bunch-by-bunch damper, coupled-bunch wakes, or a halo of the same bunch. Here a question may be asked, why is the halo needed for the feedback? Why cannot core particles play this role, when they move to the bunch head with their high transverse amplitudes acquired at the tail? The answer is that due to strong SC, the bunch slices are rigid, as it was discussed in the previous section. Strong SC means that all tune shifts are small compared with the SC tune shift, so intra-slice degrees of freedom cannot be excited, and thus the tail particles do not preserve their large amplitudes while moving to the bunch head; instead, they just follow the existing spacial pattern of the rigid-slice oscillations. 

However, what is impossible for the bunch's core, might work for its halo, whose SC tune shift is smaller. That is why the halo particles traveling from the tail to the head may remember their tail offsets, providing a tail-to-head feedback. At strong SC, this feedback would be small due to the low population of that halo, but, if the convective amplification is large enough, even a small feedback could be sufficient to ignite the absolute-convective instability, as it was suggested and modeled in Ref.~\cite{Burov:2018rmx}. Apparently, the same effect is responsible for the non-monotonic behavior of the wake instability threshold on the SC parameter reported by Y. Alexahin~\cite{AlexahinZermatt2}. A good agreement of his analytically calculated highly convective mode with the pattern of oscillations seen by A.~Oeftiger in macroparticle simulations for the same conditions deserves to be mentioned. Alas, it was the last lifetime publication of Yuri, who suddenly passed away in a year after that~\cite{YuraMemoriam}. 

In a recent Ref.~\cite{BuffatPRAB2021}, the joint action of wake and SC forces was examined by means of a numerical circulant matrix model (CMM) applied to a bunch represented by several thin rings, from one to fifty, in the longitudinal phase space with a parabolic potential well; no transverse variation of the SC tune shifts was assumed. For a single ring, i.e. for the conventional air-bag, all the conclusions of Ref.~\cite{BurovConvectivePRAB2019} with its ABS model were essentially confirmed. For many rings, a dense sequence of couplings-decouplings of the amplified radial modes were seen for the case of Gaussian longitudinal distribution, well below the air-bag TMCI threshold at the given strong SC. The authors note that "the coupling of the radial modes seems in contradiction with the results in~\cite{BalbekovPRAB2019}. Nevertheless, the model used in the latter was obtained with the strong space charge approximation, which is not met at the edge of a longitudinal Gaussian distribution. Since the CMM is not based on such an approximation, the difference can likely be attributed to that." In other words, this contradiction suggests that the longitudinal halo is an important part of the absolute-convective instability observed. The reason for this longitudinal core-halo instability could be in wider frequency range of the forces the halo particles see when they pass through the oscillating tail of the core, and this might be sufficient for them to bring the tail signal to the head. More detailed studies with this model are certainly needed for definite conclusions.     

It is clear now, that convective instabilities constitute a common obstacle for high intensity circular machines of low and medium energy, where SC is significant; they definitely take place at CERN Booster, PS and SPS rings, as well as at the Fermilab Booster. That is why it is important to understand how they behave together with other factors of beam dynamics. Transverse instabilities of a bunch with SC, wake and damper were considered in Ref.~\cite{Burov:2018zwb}. 
In Ref.~\cite{KornilovPS2013}, measurements of a microwave instability at transition crossing in PS were reported; the instability was characterized as {\it convective}. Recently an analytical model for it was proposed~\cite{Burov:2019myg} by means of Eq.~(\ref{MySSCEq}). A simple threshold formula derived there was found to be in good agreement with the data of Refs.~\cite{KornilovPS2013, MiglioratiPRAB2018}. A statement made in Ref.~\cite{KornilovPS2013} that ``The bunch parameter measurements demonstrate... that the space-charge effect does not affect the instability thresholds'' does not actually contradict to the rather weak dependence of the threshold bunch intensity $N_\mathrm{th}$ on the transverse emittance, $N_\mathrm{th} \propto \epsilon_\perp^{1/4}$ of Ref.~\cite{Burov:2019myg}, since the limited range of the emittances examined in Ref.~\cite{KornilovPS2013} and the measurement errors do not allow to resolve the rather weak dependence on the emittance on the ground of this set of measurements alone~\cite{KornilovPrivate2020}.

\section{\label{SecLong} Longitudinal Oscillations}

In this review of the SC effects in the linear longitudinal collective oscillations, we limit ourselves by bunched beams, since the coasting beam theory seems to have been completed long ago~\cite{chao1993physics, ng2006physics}. The only relatively new contribution to this theory, as far as I know, relates to the modification of the negative mass instability by the SC shift of the slippage factor, reported in Ref.~\cite{PozdeyevPRAB2009}.  

Description of bunch dynamics generally requires solving of two consecutive problems: the bunch steady state has to be found, and, after that, dynamics of its small perturbations is to be analyzed. The phase space density $F(I)$, the wake function $W(z)$, and the RF potential $U_\mathrm{rf}(z)$ have to be provided as input functions, where $I$ and $z$ are the action variable and the longitudinal position along the bunch. The steady state Hamiltonian $H(z,p)$, with $p$ as the momentum variable associated with the coordinate $z$, full potential $U(z)$, action $I(H)$ and line density $\lambda(z)$ have to be found then as solutions of self-consistent equations~\cite{Burov:2010zz}. From that, the incoherent frequency spectrum $\Omega(I)$ can be obtained as $\Omega(I)=\dd H/\dd I$. 
For SC or inductive impedances, the steady-state is determined by the distribution function $F(I)$ and the dimensionless intensity parameter
\begin{equation}
k=-\frac{2 N r_0 \eta\, \omega_\mathrm{rf}^3}{\gamma\, c\, \Omega_0^2}\, \frac{\Im Z_n}{n\, Z_0}\,.
\label{kdef}
\end{equation}
Here $N$ is the number of protons per bunch; $r_0$ is the classical radius of the beam particles; $c$ is the speed of light; $\gamma$ is the Lorentz factor; $\eta$ is the slippage factor, $\omega_\mathrm{rf}$ is the RF frequency, $\Omega_0$ is the bare RF synchrotron frequncy, $Z_n$ is the impedance at the azimuthal harmonic number $n$; for a pure SC case, $Z_n/(n Z_0) = i \ln (b/a)/\gamma^2$, where $b$ and $a$ are the aperture and the beam radii (corrections to the logarithmic factor are discussed in Ref.~\cite{Baartman:2015mta}). In particular, for a Gaussian bunch, the incoherent spectrum for the core particles at the first order of the intensity parameter $k$ is found as
\begin{equation}
\frac{\Omega(I)}{\Omega_0}= 1-\frac{k}{2\sqrt{2\pi} \sigma^3} - \frac{I}{8}\left(1 - \frac{3 k}{\sqrt{2\pi} \sigma^5} \right)\,,
\label{IncohOmega}
\end{equation}
where the bunch length $\sigma$ is measured in the RF radians, and the dimensionless units for the action $I$ are such that its maximal value in the undistorted RF bucket is $8/\pi$.     

When the steady-state problem is solved, the collective dynamics of this state can be analyzed. Usually it is done by means of linearized Jeans-Vlasov equation; the collective modes are found as its eigenfunctions. 
The main concern here relates to possible instabilities, when at least one of the eigenfrequencies has a positive imaginary part. However, one should keep in mind a possibility of convective instabilities with significant amplification along the bunch. Even if the growth rate is zero, large amplification may lead to beam loss or emittance degradation. A possibility of longitudinal convective instabilities for a bunched beam was demonstrated in Ref.~\cite{BoinePRAB2003Convective} for a barrier bucket and when both SC and resonator impedances are large enough.   

At sufficiently low beam intensity, all the collective frequencies normally supposed to lie within the incoherent spectrum, thus providing their damping by means of the Landau mechanism. As the intensity increases, some collective modes move outside the incoherent spectrum. Thus, at certain intensity, Landau damping becomes either insufficient or totally lost. The latter happens if the coherent tune is shifted by the collective interaction so far that it cannot meet resonant particles at all there; this case is termed {\it loss of Landau damping}, LLD. At the rest of this review, we concentrate on this issue, because of its high experimental importance and a recent breakthrough in this direction. Undamped collective longitudinal oscillations, interpreted as LLD-caused, were observed at RHIC~\cite{BlaskPRAB2004}, Tevatron~\cite{MoorBalb2003}, SPS~\cite{shap2006cures} and LHC~\cite{shap2011loss}.  A special role in such phenomena is played by repulsive inductance above transition energy or space charge (SC) Coulomb forces below it. Being conservative, such forces cannot drive instabilities by themselves. However, they can move the coherent frequency outside the incoherent spectrum, above the maximal incoherent frequency, in which case even a tiny wake field of the preceding bunches causes an instability. If to liken Landau damping to the beam immune system, the SC force below transition would play a role of an immunodeficiency factor, while the wake fields of other bunches would be similar to all possible viruses. In this situation, the wall inductance plays the role of a guard of Landau damping, since it behave in same way as SC, but with the opposite sign; such a guard is not necessarily sufficient, of course. Above the transition, SC and the chamber inductance switch their roles: the SC becomes the guard of Landau damping, and the inductance?the thief. Since SC normally dominates below transition, and sufficiently above it dominates the chamber inductance, the thief typically overcomes the guard. In this review, we limit ourselves by this situation of the {\it repulsive inductance}, where the force is proportional to the line density derivative, taken with the negative sign. For the vacuum chamber, this law is effective with wave numbers no higher than the inverse aperture; for hadron beams it is typically one or two orders of magnitude above the inverse bunch length. The SC impedance starts to roll off $\gamma$ times further than that.

The possibility of LLD for a bunch in the inductive vacuum chamber above transition  was first noted by F.~Sacherer~\cite{SachererLLD}; he evaluated the threshold number of particles per bunch, $N_\mathrm{th} \propto \sigma^5$. Later this result was essentially confirmed, up to numerical factors, by many authors~\cite{HofPedLLD, Balbekov:1991sf, BoineShuklaPRAB2005, Zimmermann:2008zza, burov2012dancing}. Some discrepancies between these results were explained by different distributions and model imperfections. The problem of the LLD threshold calculation looked to be essentially solved until a recent article of I.~Karpov, T.~Argyropoulos and E.~Shaposhnikova~\cite{KarpovPRAB2020} demonstrated that all the previous results were, in fact, incorrect: actually, there is no LLD threshold for such impedance, and the previous claims were all based in the insufficiency of the accepted limits on wave numbers $q$ of the perturbations or insufficient number of the mesh points. This conclusion was demonstrated in several independent ways, leaving no doubt of its correctness. It was also shown that if the inductance $i Z(q)/q$ rolls off at certain wave number $q=q_c$, the LLD threshold would be inversely proportional to that value, $N_\mathrm{th} \propto \sigma^4/q_c$. When the intensity increased, emergence of a second mode of the discrete spectrum was demonstrated. The results of this breakthrough article were soon explained and generalized in Ref~\cite{BurovPRAB2021WeakSC}. Below we reproduce this explanation.     

Qualitatively, the leading collective mode can be approximated as oscillations of a central part of size $a$ with a small amplitude $\tilde z \ll a$. Such oscillations result in the phase space density perturbation $f \simeq F'\,a\, \tilde z$ and the line density perturbation $\rho \simeq f\, a \simeq F' a^2\, \tilde z$, where $F'=\dd F/\dd I$ at zero action, $I=0$. For the inductive impedance, the related collective force is $E \simeq k \rho/a \simeq k F' a \tilde z$. For the case under study, $k>0$, this corresponds to an additional focusing seen by the collective mode, taking it above the incoherent spectrum. To avoid confusion, let us note that at the same time the incoherent frequencies are depressed by this wake. The related coherent tune shift is thus estimated as $\Delta \omega \simeq -k F'\,a/2 $, in the units of $\Omega_0$. For the same central cluster of particles, the incoherent tune spread is $\delta \Omega \simeq \pm |\Omega'|a^2/4 $, where $\Omega'=\dd \Omega/\dd I$ at $I=0$, in the total potential well. The coherent motion dominates over the incoherent as soon as $\Dom \geq \delta \Omega$, or 
\begin{equation}
a \leq 2k |F'/\Omega'| \equiv \alpha \,.
\label{alphadef}
\end{equation}
Thus, all the perturbations shorter than $\alpha$ are of relatively small incoherent tune spread; in other words, there is no Landau damping for them. The maximal coherent tune shift for them is $\Dom \simeq -kF'\alpha/2 \simeq k^2 F'^2/|\Omega'|$. The suggested estimates assume {\it a weak space charge}, when the mode size is small, $\alpha < \sigma$, leading to 
\begin{equation}
k \leq 0.2\, \sigma^5 \,.
\label{WeakCrit}
\end{equation}
Note that under this assumption the relative depression of the tune derivative $|\dd \Omega/\dd I|$~(\ref{IncohOmega}) is not large. 

The weak SC condition~(\ref{WeakCrit}) can be compared with the condition of separated multipoles, or weak headtail condition, requiring for the relative tune shift to be small. Thanks to Eq.~(\ref{IncohOmega}), the latter can be presented as
\begin{equation}
k \ll 2 \sqrt{2\pi} \, \sigma^3 \,.
\label{WeakHT}
\end{equation}

Thus, we come to a conclusion that there are four SC regimes: 
\begin{itemize}
    \item {insignificant, $\alpha \leq q_c^{-1}$ or $k<(\pi/8) \sigma^4 q_c^{-1}$;}
    \item {weak, $(\pi/8) \sigma^4 q_c^{-1} \leq k \leq 0.2\, \sigma^5 $;}
    \item {medium, $0.2\, \sigma^5 < k \ll  2 \sqrt{2\pi} \, \sigma^3$;}
    \item {and strong, $k \simeq 2 \sqrt{2\pi} \, \sigma^3$.}
\end{itemize}

For the first of them, the one  with insignificant SC, all the modes are Landau damped due to the impedance roll-off or due to the intrabeam scattering at wave numbers $q>q_c$. For the second, with weak SC, there is at least one discrete undamped mode, associated with oscillations of the relatively small central portion of the bunch, with the size $a \simeq \alpha$. This size is determined by the equilibrium between SC tune shift and nonlinearity of the RF force. For the third regime, the medium one, the RF nonlinearity already does not play a role; all bunch particles are effectively involved into the undamped collective oscillations, but the synchrotron multipoles are still well-separated, the coherent and incoherent tune shifts are relatively small, and the bunch length is mostly determined by the given emittance and RF potential. For the fourth regime, the one with the strong SC, the potential well is significantly flattened by the SC forces, and the bunch length is determined by this condition, $\sigma \simeq k^{1/3}$~\cite{Burov:1986dr, Nagaitsev:1997yq}. For all the regimes, except one of the insignificant SC, the Landau damping is lost, and even a tiny coupled-bunch (CB) wake may drive an instability.


The demonstrated back-of-the-envelope estimations essentially explain the findings of Ref.~\cite{KarpovPRAB2020} that there is no LLD threshold for the pure inductive repulsive impedance, being in agreement with the leading mode character seen in that reference. It is also instructive to note that for the attractive wake, $k<0$, the collective modes are shifted down with respect to the incoherent ones. Thus, LLD can happen only for the modes mostly associated with the high-amplitude particles, not the central ones. At high amplitudes, however, the derivative of the phase space density $F'(I)$ normally tends to zero; thus, the nonzero LLD threshold has to be expected there, which also agrees with the analysis of Ref.~\cite{KarpovPRAB2020}.   

It is worth noting that the method of estimations described in this section can be applied to any impedance. To show that, let us assume a repulsive impedance $Z(q)= \zeta (-i q)^\kappa$, with some constant parameters $\zeta$ and $\kappa$. With that, the intensity parameter $k$ is modified by a substitution $-\Im Z/q \rightarrow \zeta$, yielding the coherent tune shift $\Delta \omega \simeq -k F'\,a^{2-\kappa}/2$. Compared with the incoherent tune spread $\delta \Omega = \pm |\Omega'|a^2/4$, it leads to a conclusion that at $\kappa>0$ all the central perturbations with $a \leq |2 k F'/\Omega'|^{1/\kappa}$ are not damped. Thus, for all such impedances with $\kappa>0$ the spectral properties should be qualitatively the same as for the inductive impedance. In particular, they all must correspond to zero LLD threshold above transition. Note that the resistive wall impedance belongs to this class; it is a case of $\kappa=1/2$.    

For the SC impedance below transition, a special interest is presented by the weak SC, in terms of the suggested classification, since it relates to lowest intensities where the stability problems may already appear. In this case, Jeans-Vlasov dynamic equation reduces, after a proper scaling transformation, to a parameter-less Hermitian integral equation for the eigenfunctions and eigenvalues. The eigenvalues which lie above the upper bound of the incoherent spectrum, constitute a discrete spectrum, relating to the modes without Landau damping. The tune of the leading mode sits at $\Delta \omega = 0.22 |\Omega'| \alpha^2$ above the maximal incoherent frequency $\Omega(0)$. Formally speaking, the number of such modes is infinite; the eigenvalues have their limit point at $\Omega(0)$. When the impedance roll-ff or intra-beam scattering~\cite{BlaskPRAB2004} is taken into account, it determines the highest-frequency discrete mode. In numerical computations, the number of discrete modes is also limited by the mesh size. 

In case the inter-bunch wakes are smooth within a single bunch, the coupled-bunch growth rates can be either analytically expressed or found by means of the stability diagram~\cite{SachererLLD}, as it is shown in Ref.~\cite{BurovPRAB2021WeakSC}. Also demonstrated there is how all these results are generalized for arbitrary multipolarity of the single-bunch modes.  

%
%

%

%

\begin{acknowledgments}

%
I express my deep gratitude to the memory of my long-term friend and wise colleague Yuri Alexahin (1948-2020), discussions with whom shed light on many issues mentioned here.   

Fermilab is operated by Fermi Research Alliance, LLC under Contract No. DE-AC02-07CH11359 with the United States Department of Energy.

\end{acknowledgments}

\bibliography{bibfile}			
\end{document}